\documentclass[aps,twocolumn,prd,superscriptaddress,nofootinbib,showpacs,letterpaper,eqsecnum]{revtex4}

\usepackage{amsmath,amssymb}
\usepackage[dvips]{graphicx}
\usepackage[export]{adjustbox}
\usepackage{longtable}
\usepackage[usenames,dvipsnames]{xcolor}
\usepackage[normalem]{ulem}
\usepackage{bm}
\usepackage{comment}
\usepackage{xcolor}
\usepackage{soul}
\usepackage{hyperref}
\usepackage[percent]{overpic}
\usepackage{overpic}

\begin{document}

\newcommand{\bea}{\begin{eqnarray}}
\newcommand{\eea}{\end{eqnarray}}
\newcommand{\E}{\mathrm{E}}
\newcommand{\Var}{\mathrm{Var}}
\newcommand{\bra}[1]{\langle #1|}
\newcommand{\ket}[1]{|#1\rangle}
\newcommand{\braket}[2]{\langle #1|#2 \rangle}
\newcommand{\mean}[2]{\langle #1 #2 \rangle}
\newcommand{\be}{\begin{equation}}
\newcommand{\ee}{\end{equation}}	
\newcommand{\ba}{\begin{eqnarray}}
\newcommand{\ea}{\end{eqnarray}}
\newcommand{\SD}[1]{{\color{magenta}#1}}
\newcommand{\rem}[1]{{\sout{#1}}}
\newcommand{\alert}[1]{\textbf{\color{red} \uwave{#1}}}
\newcommand{\Y}[1]{\textcolor{blue}{#1}}
\newcommand{\R}[1]{\textcolor{red}{#1}}
\newcommand{\B}[1]{\textcolor{black}{#1}}
\newcommand{\C}[1]{\textcolor{cyan}{#1}}
\newcommand{\db}{\color{darkblue}}
\newcommand{\huan}[1]{\textcolor{cyan}{#1}}
\newcommand{\fan}[1]{\textcolor{blue}{#1}}
\newcommand{\ac}[1]{\textcolor{cyan}{\sout{#1}}}
\newcommand{\intinfty}{\int_{-\infty}^{\infty}\!}
\newcommand{\Tr}{\mathop{\rm Tr}\nolimits}
\newcommand{\const}{\mathop{\rm const}\nolimits}

\title{Magnetosphere of a Kerr black hole
immersed in magnetized plasma \\ and its perturbative mode structure} 
\author{Huan Yang}
\affiliation{Perimeter Institute for Theoretical Physics, Waterloo, Ontario N2L 2Y5, Canada}
\affiliation{Institute for Quantum Computing, University of Waterloo, Waterloo, Ontario N2L3G1, Canada}
\email{hyang@perimeterinstitute.ca}
\author{Fan Zhang}
\affiliation{Center for Cosmology and Gravitational Wave, Department of Astronomy, Beijing Normal University, Beijing 100875, China}
\affiliation{\mbox{Department of Physics, West Virginia University, PO Box 6315, Morgantown, WV 26506, USA}}
\email{fnzhang@bnu.edu.cn}
\author{Luis Lehner}
\affiliation{Perimeter Institute for Theoretical Physics, Waterloo, Ontario N2L 2Y5, Canada}
\affiliation{CIFAR, Cosmology \& Gravity Program, Toronto, ON M5G 1Z8, Canada}
\email{llehner@perimeterinstitute.ca}

\begin{abstract}
This work studies jet-like electromagnetic configurations surrounding a slowly-spinning black-hole immersed in a uniformly magnetized force-free plasma. In the first part of this work, we present a family of stationary solutions that are jet-capable. While these solutions all satisfy the force-free equations and the appropriate boundary conditions, our numerical experiments show a unique relaxed state starting from different initial data, and so one member of the family is likely preferred over the others. In the second part of this work, we analyze the perturbations of this family of jet-like solutions, and show that the perturbative modes exhibit a similar split into the trapped and traveling categories previously found for perturbed Blandford-Znajek solutions. In the eikonal limit, the trapped modes can be identified with the fast magnetosonic waves in the force-free plasma and the traveling waves are essentially the Alfv\'en waves. 
Moreover, within the scope of our analysis, we have not seen signs of unstable modes at the light-crossing timescale of the system, within which the numerical relaxation process occurs. This observation disfavors mode instability as the selection mechanism for picking out a preferred solution. Consequently, our analytical study is unable to definitively select a particular solution out of the family to serve as the aforementioned preferred final state. This remains an interesting open problem.
\end{abstract}


\maketitle

\section{Introduction}

The potential role of magnetospheres around compact objects in helping power energetic phenomena
has long been recognized~\cite{LyndenBell:1969yx,Goldreich:1969sb,1977MNRAS.179..433B}. Examples of
such phenomena include pulsars and active galactic nuclei (AGN) which represent exciting laboratories to understand
physics in extreme regimes. To date, significant insights into phenomena tied to magnetospheres have been
gained through combined theoretical (analytical and numerical) efforts and contrasting with observations. 
On the theoretical fronts, considerable efforts have concentrated on the details of the magnetosphere dynamic as it interacts with
compact objects by making use of the {\em force-free approximation} (see, e.g.~\cite{
Contopoulos:1999ga,Komissarov:2002my,
Cho:2004nn,Uzdensky:2004qu,
Asano:2005di,
McKinney:2006sc,Timokhin:2006ur,
Spitkovsky:2006np,
Komissarov:2007rc,
Kalapotharakos:2008zc,
Neilsen:2010ax,Yu:2010bp,
Palenzuela:2010xn,
Palenzuela:2010nf,
Parfrey:2011ta,Kalapotharakos:2011db,
Lehner:2011aa,
Moesta:2011bn,
Palenzuela:2011es,
Alic:2012df,
Petri:2012cs}). 
Of particular relevance to AGNs --as well as other related phenomena like gamma ray bursts, ultra-luminous X-ray
binaries, etc--, is the question of how black holes power jets. Key insights towards answering this question have been
provided by complex simulations as the inherent complexity of the relevant equations of motion, together with particularly
extreme physical set-up (which requires dealing with a black hole), hinders analytical work. Beyond particular details, 
the common message from simulations is that
the magnetosphere can tap into the black hole's kinetic energy reservoir and induce a strong Poynting flux from the magnetosphere. This understanding, complemented with
further studies as to the stability of jets, interaction with
a possible accretion disk, etc. has been instrumental in putting forward particular models of how jets may be 
launched (see e.g.~\cite{Meier:1481607,Romero:2014uma} and references cited therein).

Despite the aforementioned advances, it is arguably desirable to also gain access to black hole-magnetosphere interaction
through analytical means. This would allow a more convenient way to explore questions like:
What collimates the emitted radiation? What is the electromagnetic field configuration within the jet? How do the electromagnetic and plasma perturbations propagate within the jets? These questions are ultimately intimatedly tied to observations.

To help answer these questions, we here carry out analytical studies complemented by numerical investigations 
within the force-free approximation and study the corresponding magnetosphere as it interacts with a (slowly) spinning
black hole. That such approximation is relevant in this context was pointed out over four decades ago in\cite{LyndenBell:1969yx}.
At the core of this observation is the fact that a particle-production cascade will take place and the magnetosphere
will be composed of a tenuous plasma which will reach equilibrium when the electric field becomes orthogonal
to the magnetic field. The electromagnetic energy density of the plasma dominates the plasma mass density \cite{1977MNRAS.179..433B}, 
and it is expected to remain so over time; correspondingly the transfer of energy and momentum into the plasma is negligible.
In mathematical terms, 

\begin{equation}\label{eqffe}
F_{\mu\nu} j^\nu=0\,,
\end{equation}
where $F_{\mu\nu}$ is the electromagnetic field tensor and $j_\nu$ is the $4$-current density. This is the so
called {\em force-free condition} 
\cite{1977MNRAS.179..433B,Goldreich:1969sb,Gruzinov:2007se} which results from $\nabla_a T^{ab}_{EM} = 0$.

In the present work, we focus particularly on revisiting the question of jets induced by spinning black holes. 
To simplify the analysis, we shall assume that the black hole region is threaded by an electromagnetic field configuration
which, asymptotically, is purely magnetic and parallel to the black hole's angular momentum. Consequently,
the system has a rotational symmetry around the spin axis. In addition, we restrict ourselves to slowly-rotating black holes, and solve
the force-free equations through a perturbative expansion (considerably generalizing the solutions found
in~\cite{1977MNRAS.179..433B,McKinney:2004ka}). 
Despite these approximations, we expect the physics discussed here to carry through to more general cases, for 
example with misaligned spins \cite{Gralla:2015wva} or more rapidly spinning black holes,
and expectectations already supported by numerical simulations (e.g.~\cite{McKinney:2004ka,Palenzuela:2010xn}).
We note that while prior analytical efforts have also concentrated in our case of interest~\cite{McKinney:2004ka},
we here adopt the powerful formalism of Euler potentials and the techniques of exterior calculus as discussed in 
Refs.~\cite{Carter1979,1997MNRAS.286..931U,1997MNRAS.291..125U,
1997PhRvE..56.2181U,1997PhRvE..56.2198U,1998MNRAS.297..315U,Gralla:2014yja} and so can explore the possible solutions
in broader terms. Our analysis
will reveal the existence of a family of solutions describing jets which emanate from the black hole and propagate towards infinity along the symmetry axis. These jets are enclosed by a tube-like surface, within which energy and currents flow to infinity, and the electromagnetic field does not asymptote to the external constant field set up initially. 
Outside of this tube, there is no charge, current or energy flux, and the field matches the external driver at infinity.
An important observation is that   
all solutions within this family exhibit the same qualitative behavior we have just described, and we have not found a definitive selection criteria as to which particular one is more physically realistic. 
In particular, none of  these solutions appear to have an instability over the light-crossing timescale (across
the tube), 
as suggested by the numerical investigation in Sec. \ref{sec3}.
However,  numerical studies of the jet solutions obtained with different initial data does support the existence
of a single relaxed state determined only by the black hole and the external field. Thus, 
it appears that such a criteria may very well exist and we are  failing to impose it 
when solving for our stationary jet-like solutions.
From our discussion, it will be clear that the task of selecting a particular solution is tightly connected with 
the resulting angular velocity of magnetic field lines and the development of a current sheet --where the force-free
equations are inconsistent without removing some amount of energy--. These are rather subtle and long-standing issues in the search for FFE solutions. We expect any successful future resolution of them to also provide important clarity to FFE dynamics in general.

Having obtained such a family, we then investigate its perturbative mode structure. 
We identify a clean and simple separation into ``trapped modes" and ``traveling waves",  
which are analogous to the findings made on the perturbed Blandford-Znajek configuration \cite{Yang:2014zva}.
Interestingly, the trapped modes behave similarly to the vacuum electromagnetic quasinormal modes (QNM) of black holes \cite{Teukolsky,Teukolsky:1972my,ReggeWheeler1957,Zerilli1970b,
Berti2009,Yang:2013shb}. In fact, in the short wavelength (eikonal) limit, these trapped modes become the fast-magnetosonic waves of the force-free plasma and satisfy exactly the same equations that the vacuum QNMs
obey in the same limit. On the other hand, the traveling waves generically carry charge, and propagate inward towards the black hole or outward towards spatial infinity along the background magnetic field lines. In the eikonal limit, they become the Alfv\'en waves of the force-free plasma. These observations are consistent with the analysis of Uchida \cite{1997MNRAS.291..125U}, where he finds fast magnetosonic waves and Alfv\'en waves by examining local dispersion relations. Moreover, our study reveals no mode instability at the light-crossing timescale for any of these jet-like solutions, in contrast with the fast-variation that we observe numerically in Sec.~\ref{sec3}. This suggests that stability can not be the criteria to
single out the unique relaxed state observed in numerical simulations.

We  organize the paper  as follows. In Sec.~\ref{sec2}, we introduce the family of jet-like solutions and discuss their validity. In Sec.~\ref{sec3}, we perform our numerical studies with different initial data. The final relaxed states are extracted and compared with each other. In Sec.~\ref{sec4} we apply the formalism developed in Ref.~\cite{Yang:2014zva} to analyze perturbations of these jet-like configurations, where discussions on mode instability are made. Finally, we conclude in Sec.~\ref{sec5}. Throughout this manuscript we adopt the geometric units, setting the gravitational constant $G$ and the speed of light $c$ to one.

\section{Stationary jet solutions}\label{sec2}

In order to solve for the stationary field configuration in the presence of a slowly-spinning black hole, we start with a background configuration (the ``Wald-type solution") for a Schwarzschild black hole, and then compute the changes required after a spin is introduced. We shall present a family of solutions which satisfy the force-free equations and the boundary conditions. 

\subsection{The Wald solution for Schwarzschild black holes}

The Wald solution \cite{PhysRevD.10.1680} describes the electromagnetic field distribution surrounding a  axis-symmetric black hole immersed in a constant magnetic field. In our case, we start with the Wald configuration for a Schwarzschild black hole. For simplicity, let us assume that the magnetic field is oriented along the $z$ direction (while $x, y$ directions are orthogonal to $z$), with its strength being $B_0$. More specifically, The magnetic field in the Wald solution is (in orthonormal coordinate)

\begin{equation}
{\bf B} = B_0( \cos\theta {\bf e}_{\hat r} -\sqrt{f} \sin\theta {\bf e}_{\hat \theta})\,,
\end{equation}
with $f \equiv 1-2M/r$. The corresponding field tensor is
\begin{align}\label{eqbg}
F = & B_{\hat \phi} d \hat{r} \wedge d \hat \theta -B_{\hat \theta} d \hat r \wedge d \hat \phi+B_{\hat r} d \hat \theta \wedge d \hat \phi \nonumber \\
=& \frac{B_0}{2} d (r \sin\theta)^2 \wedge d \phi \nonumber \\
=& B_0 d (r \sin\theta \cos\phi) \wedge d (r \sin \theta \sin\phi) \nonumber \\
 \equiv & B_0 dx \wedge dy\,.
\end{align}

In the language of Refs.~\cite{Carter1979,1997MNRAS.286..931U,1997MNRAS.291..125U,
1997PhRvE..56.2181U,1997PhRvE..56.2198U,
1998MNRAS.297..315U,Gralla:2014yja}, 
the field tensor can be expressed in terms of the differentials of two ``Euler potentials'', in the form of ${ F}=d \phi_1 \wedge d \phi_2$, and these potentials are not unique. With the Wald solution, 
we can either choose the gauge in which the Euler potentials are $\psi_1 \equiv r \sin\theta \cos \phi$ and $\psi_2 \equiv r \sin \theta \sin\phi$, or 
\bea \label{eq:WaldPot}
\phi_1 =r^2 \sin^2\theta/2\,, \quad \phi_2 =\phi\,,
\eea
together with a multiplicative constant $B_0$. We will choose the latter in our calculations in the next section. Recall
that the Wald solution was originally derived for vacuum (without plasma) Maxwell equations but in the case of a non-spinning black hole it satisfies all the force-free requirements as there is no current, charge, or electric field in the spacetime. Therefore in the non-spinning case, 
it also provides a solution for a Schwarzschild black hole immersed in constantly magnetized force-free plasma.

\begin{figure}[t,b]\centering
\includegraphics[width=0.65\columnwidth]{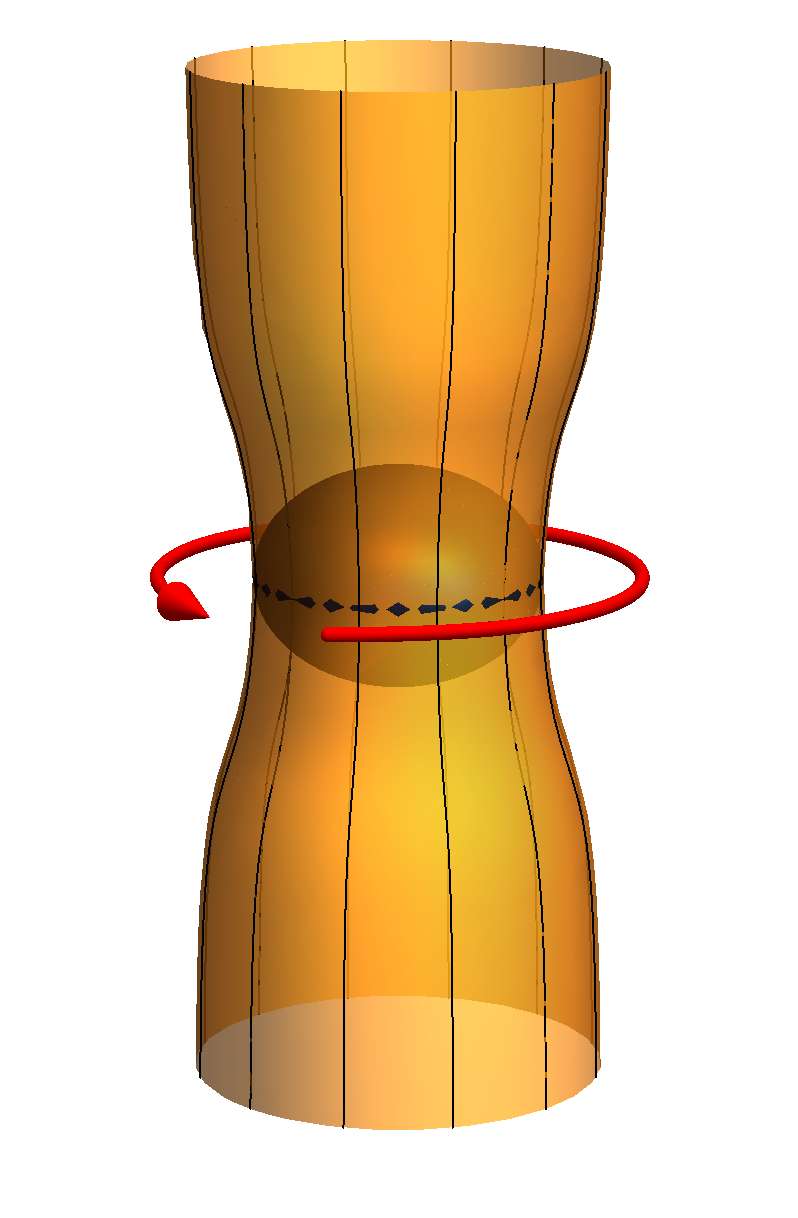}
\caption{A schematic illustration for a jet-like solution launched from a rotating black hole. The magnet field lines originated at the $\theta=\pi/2$ ring of the horizon separate inner and outer regions of the jet. In the inner region, current and the angular velocity of the field lines are related by the Znajek boundary condition, whereas they are less constrained in the outer region. The red arrow labels the rotation direction of the black hole.}
\label{fig:tube}
\end{figure}
\subsection{Small spin expansion and jet solutions}

As shown in Ref.~\cite{Gralla:2014yja}, the force-free Euler potentials of a stationary, axis-symmetric spacetime can only take on a very restrictive form
\begin{equation}\label{equchida}
\phi_1 = \psi(r ,\theta), \quad \phi_2 = \psi_2(r, \theta)+\phi-\Omega_F(\psi) t\,,
\end{equation}
where $\psi$ corresponds to the magnetic flux enclosed by the circle at constant $r$ and $\theta$, and $\Omega_F$ is the angular velocity of the field lines passing through that circle. In addition, the force free condition Eq.~\eqref{eqffe} turns into the pair of equations:
\bea \label{eq:FFEqs}
d\phi_1 \wedge d* F = 0\,, \quad 
d\phi_2 \wedge d* F = 0 \,.
\eea

The potentials \eqref{eq:WaldPot} are obviously in a form consistent with Eq.~\eqref{equchida}, and we will use them as the background configuration.
We can then turn on the black hole spin and investigate the change of field distributions. As the spin of black hole $a$ is assumed to be small, the deviation of the Euler potentials from their background values may be solved in a series expansion of $a$. Since $\phi_1$ and $\phi_2$ are given by Eq.~\eqref{eq:WaldPot} in the non-rotational limit, we shall have the following expansions
\begin{align}\label{eqexpanv}
&\psi= r^2\sin^2\theta/2 + a\, \psi^{(1)}(r, \theta) +a^2 \, \psi^{(2)}(r, \theta)+\mathcal{O}(a^3)\,, \nonumber\\
&\psi_2 = a\, \psi^{(1)}_2(r, \theta) +\mathcal{O}(a^2)\,, \nonumber \\
& \Omega_F= a\, \Omega_F^{(1)}(\psi)+\mathcal{O}(a^2)\,.
\end{align} 
We expand $\psi$ to higher orders in $a$ here for reasons which will become clear later. In addition, the polar current is defined as
\begin{equation} \label{eq:PolarCurr}
*(d \psi \wedge d \psi_2) = \frac{I}{2 \pi} dt \wedge d \phi\,,
\end{equation}
where the operation with Hodge star $*$ in a slowly-spinning Kerr spacetime is discussed in the Appendix of Ref.~\cite{Yang:2014zva}. Using the first expression of Eq.~\eqref{eq:FFEqs}, this polar current can be shown to be a function of $\psi$ only \cite{Gralla:2014yja}, as the current flows along the magnetic field lines in the frame that co-rotate with the field lines. Moreover, the polar current can also be expanded in $a$: $ I = a I^{(1)}(\psi)+\mathcal{O}(a^2 )$, and we have 
 \begin{equation}
a * [d (r^2 \sin^2\theta) \wedge d \psi^{(1)}_2] = \frac{a I^{(1)}}{2 \pi} dt \wedge d \phi\,,
  \end{equation}
and so 
\begin{align} \label{eq:psi2vsI}
d \psi \wedge d \psi_2 &= -**(d \psi \wedge d \psi_2) \notag \\ &\approx \frac{a I^{(1)}}{2 \pi} \frac{1}{f \, \sin\theta} d\theta \wedge dr\,. 
\end{align}

The polar current $I$ and the angular velocity $\Omega_F$ are not independent of each other -- they are related by the Znajek boundary condition \cite{Znajek1977, Beskin2000, Komissarov2004}, which requires that the field on the horizon be regular. In the current context, Znajek boundary condition reads 
\begin{align}\label{eqzna}
\left . a I^{(1)} \right |_{r_+} &= \left .2\pi(a\, \Omega^{(1)}_F-\Omega_H) \psi_{, \theta} \sin \theta \right |_{r_+} \nonumber \\
&\approx 2\pi (a\, \Omega^{(1)}_F-\Omega_H)\rho^2
\frac{\sqrt{r^2_+-\rho^2}}{r_+}\,,
\end{align} 
where we have defined $\rho \equiv r \sin \theta$ as the cylindrical radial coordinate, and $r_+ \equiv M+\sqrt{M^2-a^2}$ is the outer horizon radius and $\Omega_H \equiv a/(2 M r_+)$ is the horizon angular velocity. As $I^{(1)}$ and $\Omega^{(1)}_F$ are both function of $\psi \approx \rho^2/2$, they must be approximately functions of $\rho$ as well. Consequently
their functional relationship in the entire region where $\rho \le r_+$ is the same as their relationship on the horizon. It is then clear that the Znajek boundary condition has naturally divided the spacetime into two parts (see Fig.~\ref{fig:tube} for a schematic illustration). The inner part is enclosed by a tube, where the polar current and the field-line angular velocity is related with each other by Eq.~\eqref{eqzna}. Outside of the tube these two quantities are less constrained. In principle, any solution of the form prescribed by Eq.~\ref{equchida} that satisfies the stream equation (i.e. the second force-free equation in \ref{eq:FFEqs}, which we haven't used. See next section) is admissible.

There are  arguments favoring zero $\Omega_F$ and polar current density in the outer region: In the inner region, the rotation of the field lines and the currents are both sourced by the rotation of the black hole, whereas in the outer region they have to be driven by sources at infinity. As it is not physically motivated to have such sources at infinity, we should expect $\Omega_F=0$ and $I=0$. In fact, as shown in \cite{Gralla:2014yja}, we have
\begin{equation}\label{eqelflux}
\frac{d \mathcal L}{d t} = - \int_{P} I d \psi\,,\quad \frac{d \mathcal E}{d t} = -\int _{P} I \Omega_F d \psi\,,
\end{equation}
where $P$ identifies some curve in the $(r, \theta)$ plane. Physically, the equations above imply that 
non-vanishing $\Omega_F$ and $I$ would
always introduce energy and angular momentum fluxes, which are injected (along the field lines) by the sources at infinity and eventually escape out. As it is physically more nature to imposes outgoing boundary conditions at infinity, such flux injections should not be allowed. While these arguments, strickly speaking, only apply to regions near the outer boundary, instead of the entire space outside of the tube,  previous numerical simulations do favor zero $\Omega_F$ and constant $I$ outside the jet-tube \cite{Palenzuela:2010xn}. Therefore in the analysis below we mainly focus on the solution inside the tube. 

We note however, that there is one subtlety when we allow the presence of a current sheet, where currents can be launched at, and move away from the equatorial plane. This may provide an additional driving mechanism for $\Omega_F$ and $I$ outside of the black-hole-driven tube, and effectively enlarge the jet-tube. This is likely what we observe in Fig.~\ref{fig:numerics_late}, where the tube radius of the final relaxed state is clearly greater than $r_+$. As the analytical description of the dynamical effect of a current sheet feeding back onto the force-free plasma remains elusive, we are not able to further constrain these current-sheet-driven regions.

Finally, we note that although the discussion in this section relies on an expansion in $a$, we expect this basic picture of a tube dividing the magnetosphere into a jet region and an outside region to remain valid when $a$ is not a small quantity. In particular, inside the jet, the Znajek boundary condition (without taking the small $a$ limit as in Eq.~\ref{eqzna}) should still provide us with a relationship between $I$ and $\Omega_F$. 

\subsection{The stream equation}

The stream equation is derived from the force-free conditions (Eq.~\ref{eqffe} or more precisely the second expression in Eq.~\eqref{eq:FFEqs} \cite{Gralla:2014yja}), and it is also often referred to as the Grad-Shrafranov equation \cite{GradRubin,Shafranov}. In a covariant language it can be written as \cite{Gralla:2014yja}
\begin{equation}
\nabla_a (|\eta |^2 \nabla^a \psi)+\Omega_{F\,,\psi} \langle dt, \eta \rangle |d \psi|^2-\frac{I \, I_{,\psi}}{4 \pi^2 g^T}=0\,,
\end{equation}
where
\bea
\eta \equiv d\phi - \Omega_F(\psi)dt\,.
\eea

Given $I$ and $\Omega_F$'s dependence on $\psi$, the stream equation determines the distribution of $\psi$ in the spacetime. In the case of a slowly-spinning Kerr background, it has been shown in Ref.~\cite{1977MNRAS.179..433B}
\footnote{With a slightly different set of notations. We have the following identifications between the quantities of that paper and those utilized in the current presentation: $X(r, \theta) \leftrightarrow \rho^2/2$, $x(r,\theta) \leftrightarrow \psi^{(1)}$, $W (r ,\theta)\leftrightarrow \Omega^{(1)}_F$, and $Y(r,\theta) \leftrightarrow I^{(1)}/(2\pi)$.}
 that $\psi^{(1)}$ is zero and $\psi^{(2)}$ is sourced by $\Omega^{(1)}_F$ and $I^{(1)}$. In particular, for any $\Omega^{(1)}_F$ and $I^{(1)}$, as long as they satisfy the Znajek boundary condition, the resulting $\psi^{(2)}$ from the stream equation is well defined (bounded). Therefore the small-$a$-expansion analysis implies that there is a family of force-free electrodynamic solutions in the jet-tube, corresponding to different choices of the angular velocity function $\Omega_F$. As mentioned earlier, inside the jet-tube, the magnetic field does not asymptote to the external driving field, which is a fundamental difference between force-free electrodynamics and vacuum electromagnetism. In Sec. III we shall explore this observation numerically,  but first, let us discuss two physically interesting solutions.
 
 \subsection{Physically interesting solutions}\label{physical}
We have now obtained a family of regular (on the horizon and elsewhere) solutions, which is too broad for the purpose of understanding jet physics (also see \cite{Nathanail2014} for discussions on different stationary solutions). Ideally, we would like to single out a unique physically realistic solution. However, as recent experiences with analytical
searches for solutions in other spacetimes (such as in the near-extremal Kerr back holes \cite{Zhang:2014pla,Lupsasca:2014hua,Lupsasca:2014pfa}) show, being unable to pick a unique solution is a natural state of affairs. Routinely, the condition of regularity is insufficient to pin down a unique solution.

We mention in passing that a jet solution has been found in Refs.~\cite{Pan:2014bja,BeskinBook}, and their conclusion would
seem to imply the solution found is unique. 
Motivated from the $I$ vs. $\Omega_F$ relationship of the monopole solution, the jet solution obtained in these works is indeed valid and belongs to the family of solutions we have presented here. However, the additional constraint employed for
singling out a unique solution in~\cite{Pan:2014bja,BeskinBook} is motivated by the desire to link to the split monopole 
solution which need not be the physically natural condition.

Despite the lack of any solid conditions to single out particular solutions from the family, we nevertheless note that there are solutions within this family possessing interesting physical properties. For instance,
demanding that the solution maximizes the energy radiated $d \mathcal{E}/dt$ from the jet, it is clear from Eq.~\eqref{eqelflux} and the Znajek boundary condition Eq.~\eqref{eqzna} that the optimal choice is $\Omega_F = \Omega_H/2$. In this case, the field tensor inside the flux tube is given, to $\mathcal{O}(a)$, by:
\bea
F&=& \frac{a r^2 \sin \theta  \sqrt{4 M^2-\rho^2 }}{16 f M^3}d r\wedge d\theta 
\notag \\
&&
+\frac{a r^2 \sin \theta  \cos \theta   }{8 M^2}dt\wedge d\theta-\frac{a r \sin ^2\theta  }{8 M^2}dr\wedge dt
\notag \\
&&
+r^2 \sin \theta  \cos \theta  d\theta \wedge d\phi 
+r \sin ^2\theta  dr\wedge d\phi 
\eea
while outside the tube, we have the background ``Wald solution'' of Eq.~\eqref{eqbg}. 

If one instead requires that the energy stored within the tube is minimized, we realize that only the part of the tube in the asymptotically flat region needs to be considered, which accounts for most of the volume and contains most of the energy. In other words, we can simply require
\begin{align}
\rho_E &\propto B^2+E^2 =  B^2_0\left [1+a^2 \frac{{I^{(1)}}^2}{(2\pi)^2 \rho^2}+a^2{\Omega^{(1)}_F}^2 \rho^2\right ] \nonumber \\
&=B^2_0\left [1+\frac{1}{r^2_+}(\Omega_F-\Omega_H)^2 \rho^2 (r^2_+-\rho^2)+\Omega^2_F \rho^2\right ]\,,
\end{align}
to be minimized, giving
\begin{equation}
a\Omega^{(1)}_F = \frac{r^2_+-\rho^2}{2r^2_+-\rho^2} \Omega_H\,.
\end{equation}
For this case, the field tensor inside the tube is 
\bea
F &=&\frac{r \xi }{2 f M} \sqrt{4 M^2-\rho^2} dr\wedge d\theta  
\notag \\
&&
+r^2 \sin \theta  \cos \theta  d\theta \wedge d\phi +r \sin ^2\theta  dr\wedge d\phi \,,
\notag \\
&&
+  r \cos\theta \chi dt\wedge d\theta 
- \sin\theta \chi dr\wedge dt 
\eea
where
\bea
\chi = \xi \left(1
-\frac{ \rho^2  }{4 M^2}\right), \quad
\xi = \frac{a \rho }{8 M^2-\rho^2}
\eea

\section{Numerical searches for the relaxed state}\label{sec3}

A powerful way to analyze the steady state solutions possibly achieved by a system described by the force-free equations
is to study numerical solutions obtained under different initial configurations and examine their late-time properties.
To this end, we employ a numerical code described in~\cite{Palenzuela:2010xn} 
which has been thoroughly tested and employed to study different 
systems in e.g~\cite{2010Sci...329..927P,2011CQGra..28m4007P,2011PNAS..10812641N}. This code implements the force-free equations
and ensures sufficient resolution is achieved efficiently through
the use of adaptive mesh refinement (AMR) via the HAD  computational infrastructure
that provides distributed, Berger-Oliger
style AMR~\cite{had_webpage,Liebling} with full sub-cycling
in time, together with an improved treatment of artificial boundaries~\cite{Lehner:2005vc}. We also note
that a Cartesian grid stucture is employed without imposing any symmetries. As a result, truncation errors introduce
the typical $m=4$ perturbation. As we will illustrate next, a rather unique steady state solution is achieved irrespective
of the initial configuration, provided the asymptotic conditions on the electromangetic field are the same. 
This observation, together with the fact that the $m=4$ perturbation does not seem to affect the observed behavior 
reinforces the belief that it is attractive and stable.

For simplicity, and since the electromagnetic field's back-reaction on the spacetime is negligible, we adopt
for the geometry a spinning black hole solution in terms of ingoing Eddington-Finkelstein coordinates. For the initial electromagnetic
fields, we adopt the two physical cases described
in section \ref{physical} which we label case P1 and case P2 for the radiated energy maximizing and minimizing energy stored configurations
respectively. As well, we use an additional initial data with $E_a = F_{ab} n^b = 0$ and $B_a = *F_{ab} n^b = B_0 \delta_a^z$ (case A1).

\begin{figure}[t,b]\centering
\includegraphics[width=0.85\columnwidth]{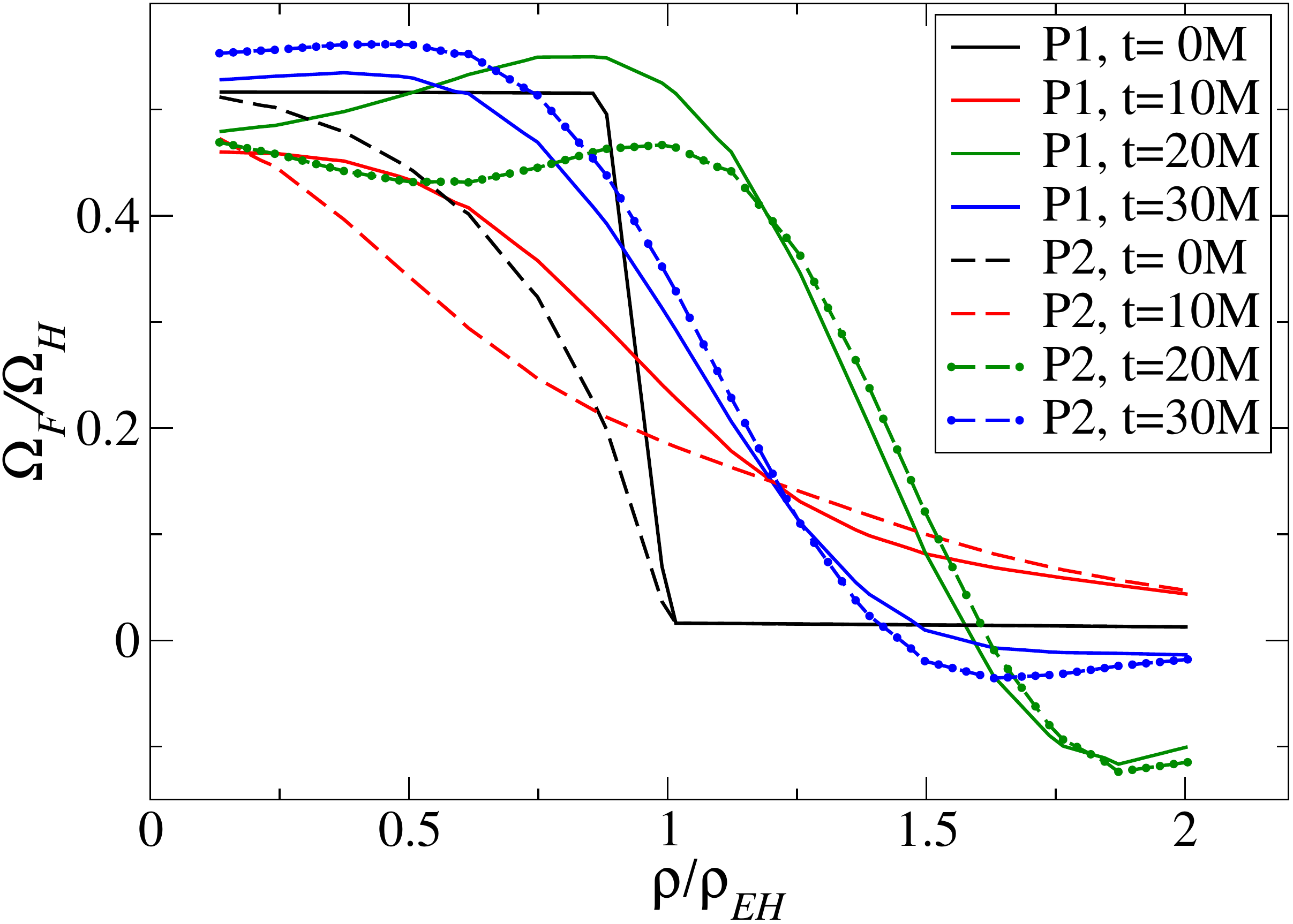}
\caption{$\Omega_F$ versus x at $z=7 M$ at different early times in a spinning
black hole with $a/M=0.1$ for cases P1, and P2.}
\label{fig:numerics_early}
\end{figure}

\begin{figure}[t,b]\centering
\includegraphics[width=0.85\columnwidth]{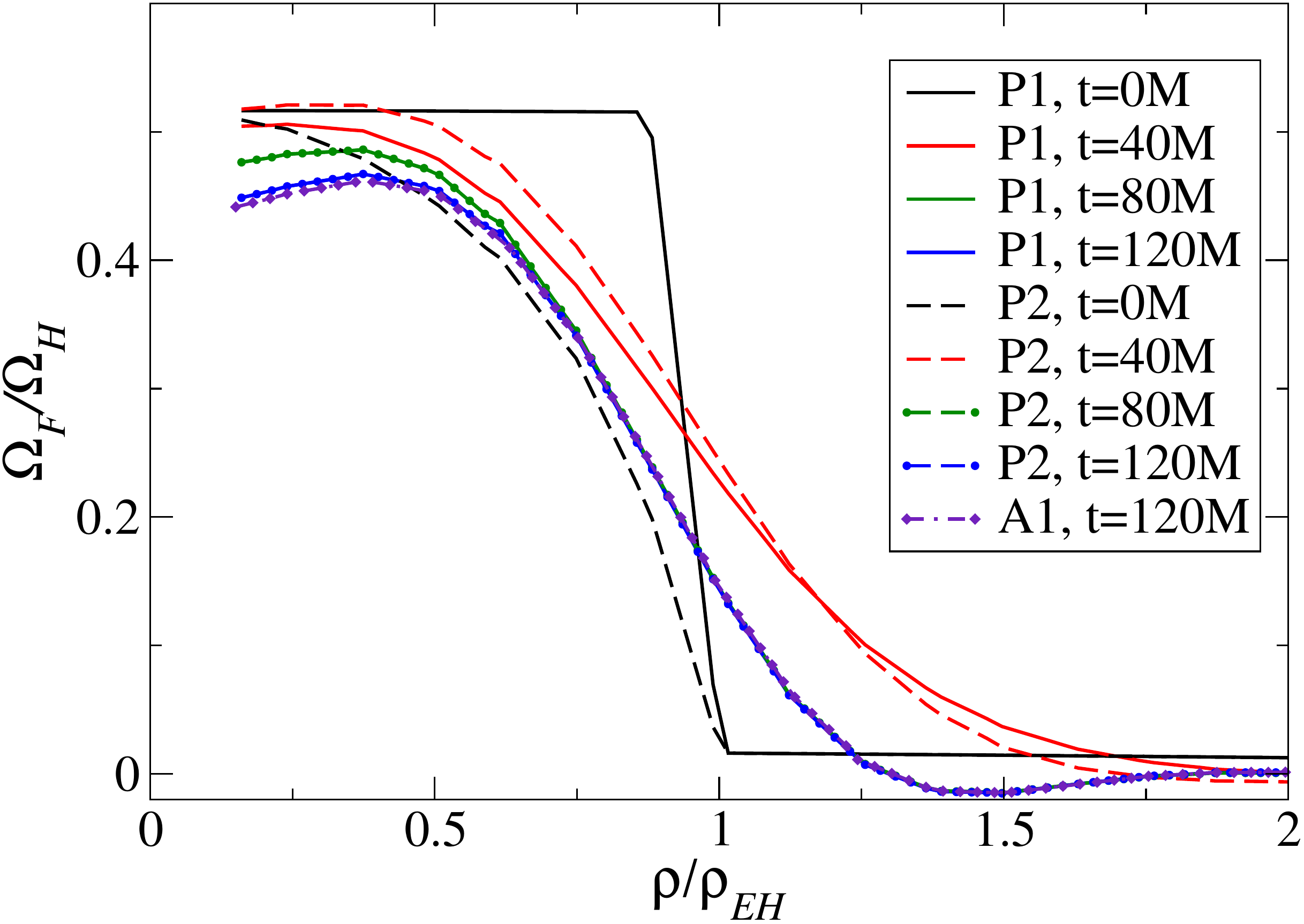}
\caption{$\Omega_F$ versus x at $z=7 M$ at different times in a spinning
black hole with $a/M=0.1$ for cases P1, and P2 and also the case A1 at $t=120M$.
Despite the initial differences, all three solutions relax to a single one.}
\label{fig:numerics_late}
\end{figure}

\begin{figure}[t,b]\centering
\includegraphics[width=0.85\columnwidth]{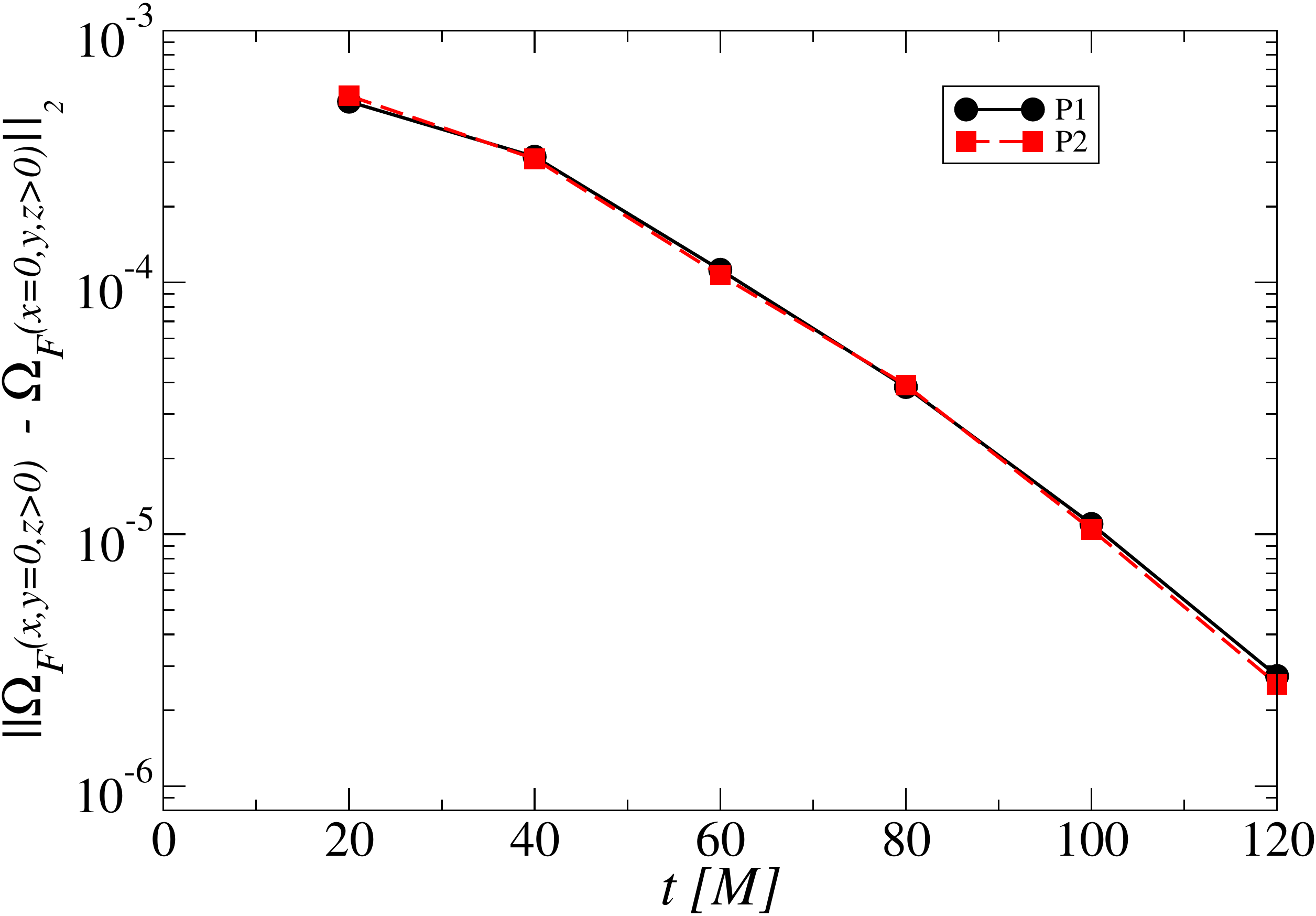}
\caption{$L_2$ norm of the difference between the vaues of  $\Omega_F$ obtained with the
two ``physical'' cases for a spinning black hole with $a/M=0.1$.}
\label{fig:numerics_diffID}
\end{figure}

We note that P1, P2 of these initial configurations are stationary solutions of the force-free equations, while A1 is not. Nevertheless, in all cases, one observes an Alfv\'en-wave-like transient
that is radiated away from the equatorial plane and then, after a few crossing times, the solution is seen in all cases to relax to the same solution.
As an illustration, Fig.~\ref{fig:numerics_late} shows the value of $\Omega_F$ obtained after the relaxed state is reached
for the P1 and P2 cases, together with a higher resolution run for case A1 (Fig.~\ref{fig:numerics_early} shows the earlier evolution of this quantity). Clearly, despite different initial configurations
adopted, there is an agreement in the relaxed states achieved by all cases. Moreover, the final solution is axisymmetric
as expected, which is illustrated
in Fig.~\ref{fig:numerics_diffID} that plots the (norm of the) difference between the $\Omega_F$ values on the $x=0$ and $y=0$ planes, as a function of time and for both the P1 and P2 cases. As time progresses, the differences decay exponentially to zero.

\section{The mode structure of the jet solutions}\label{sec4}

In addition to the stationary configurations, it is also interesting to study the dynamical evolution of black hole magnetospheres and the jets. In fact, many radio observations of quasars and pulsars have revealed significant time-dependent variations (see e.g.~Refs.~\cite{1984AJ.....89.1111H,
1987AJ.....94.1493H,
1989Natur.337..442Q,1995ARA&A..33..163W}), which invite further studies of the generation and propagation mechanisms of perturbations within the jets. In Ref.~\cite{Yang:2014zva} we studied the mode structure of the Blandford-Znajek split-monopole configuration, and here we apply similar techniques to the axis-symmetric and stationary jet-like solutions discussed in Sec.~\ref{sec2}.  We shall first explain the perturbation method, and demonstrate its usage by finding the modes of general axis-symmetric and stationary solutions in flat spacetime. After that we shall tackle the more difficult task of analyzing the mode structure of jet-like solutions of the slowly-rotating Kerr spacetime. At the end of the section, we will comment on the stability of this family of solutions.

\subsection{Formulation}

For any stationary and axis-symmetric solution described by Eq.~\eqref{equchida}, the perturbed force-free configurations can be written as
\begin{equation}
F = F_B+\delta F \nonumber= (d \phi_1 + \epsilon d \alpha) \wedge (d \phi_2+ \epsilon d \beta)\,,
\end{equation}
where $\alpha, \beta$ are the perturbative Euler potentials and $\epsilon$ is a flag that helps us track the order of these perturbative fields. 
The perturbative part of the field tensor is given by 
\bea
\delta F = d\alpha \wedge d \phi_2 + d\phi_1 \wedge d \beta\,.
\eea
At $\mathcal{O}(\epsilon)$, the force-free Eqs.~\eqref{eq:FFEqs} then become
\bea 
d\alpha \wedge d*F+d\phi_1 \wedge d*\delta F &=& 0\,, \label{eq:FFEq1Raw} \\
d\beta \wedge d*F+d\phi_2 \wedge d*\delta F &=& 0\,. \label{eq:FFEq2Raw} 
\eea

Before we attempt to solve these equations, we note that when we substitute in the background Euler 
potential $\phi_2$ (\ref{eq:FFEqs}) for the slowly rotating solutions, the force-free equations above pick up an 
explicit $t$ dependence tied to our gauge choice. This will of course not feed into the final results for the perturbed field tensor, which should respect the time-translational symmetry of the background solution, but is certainly undesirable. 
It is both preferable and convenient to refine the perturbing field $\beta$, so that the resulting force-free field tensor becomes
\begin{align}\label{eqwaveq}
F &= F_B+\delta F \nonumber \\
&= ( d\psi + d \alpha) \wedge (d \psi_2 -d (\Omega_F t) + d \phi+d [\beta-t \Omega'_F \alpha] )\,,\nonumber\\
\end{align}
where the perturbative Euler potentials $\alpha, \beta$ are now invariant under a time-redefinition $t \rightarrow t+C$. Under this choice the force-free equations result,
\begin{align}
&d \alpha \wedge d*F_B + d\psi \wedge d*\delta F  =0\,,\label{eq:FFEq1} \\
&( d \beta- \alpha  \Omega'_F dt )\wedge d*F_B 
\notag \\ & \quad \quad
+ (d \psi_2 -\Omega_F d t + d \phi) \wedge d*\delta F=  0\,. \label{eq:FFEq2}
\end{align}
which contain no explicit $t$ dependence. 

Note that these equations are coupled and non-separable. To proceed our analysis, we next 
adopt some simplifying approximations which enable us to handle them. These approximations
will allow us to study their modal structure, under different limits, in the next sections.

\subsection{Eikonal limit perturbations in flat spacetime}\label{sec42}
We begin with the simple case of a flat background spacetime. Despite its simplicity, it is already a 
good approximation when we examine that part of the jet that lies far away from its host black hole. In addition, we shall leave the functions $\psi_2$ and $\Omega_F$ generic, without specifying their detailed forms, but we do assume that the flux function $\psi$ is just $\rho^2/2$, which is consistent with 
 the class of solutions described in Sec.~\ref{sec2}. If one wishes to consider the jet of, say, a rapidly-spinning black hole in the far-zone in the future, it will be straightforward to insert the corresponding $\psi(\rho)$ and repeat the analysis in this section.

Note that as the perturbations are in general not stationary and/or axi-symmetric, we do not have simplifications to the perturbing Euler potentials like those leading to Eq.~\eqref{equchida}. Nevertheless, 
as the background jet solution is axis-symmetric, we can Fourier expand $\alpha$ and $\beta$ in Eq.~\eqref{eqwaveq} in $\phi$:
\bea\label{eqfdecom}
\alpha= \sum_m e^{i m \phi } \alpha_m(t, r, \theta),\quad \beta=\sum_m e^{i m \phi} \beta_m(t, r, \theta)\,,\nonumber\\
\eea
and solve for $\alpha_m, \beta_m$ for individual $m$'s separately without worrying about the background introducing coupling between different $m$'s. The resulting wave equations (c.f. Eqs.~\eqref{eq:FFEq1} and \eqref{eq:FFEq2}) are still coupled (between $\alpha$ and $\beta$) and non-separable, with the following form
\begin{align}\label{eqgwe}
H_1 \alpha_m +V_1 \beta_m =0\,,\nonumber \\
H_2 \beta_m +V_2 \alpha_m=0\,.
\end{align}
The quantities $H_{1,2}$ and $V_{1,2}$ are lengthy differential operators which we do not present until further
restrictions are introduced. In order to make further progress and obtain physical solutions, we focus on the short-wavelength (compared to the size of the black hole) perturbations, which are relevant for most astrophysical scenarios. Under this eikonal limit, one writes $\alpha_m, \beta_m$ as
\bea\label{eqwkb}
\alpha_m \sim A_m e^{i S/\epsilon},\quad \beta_m \sim B_m e^{i S/\epsilon}\,.
\eea
Here $A_m$ and $B_m$ are the amplitude functions which are assumed to vary slowly with location (on the length scale of the black hole size), while $S$ is the common phase factor shared by both wave components, and it changes much faster than $A_m$ and $B_m$ under the eikonal approximation. To keep track of this separation of scales, we introduced the book-keeping symbol $\epsilon$ which in essence labels the WKB orders. Similar WKB treatments for uncoupled waves can be found in Refs.~\cite{Yang:2012he,Yang:2012pj,Yang:2013uba,Yang:2013shb}, and readers interested in further details can consult these references. Note that the phase matching between $\alpha_m$ and $\beta_m$ follows from the assumption that we are solving for a single eigen-mode, in which case phase coherence should be preserved during propagation. In general though, the field perturbations could comprise a linear combination of a multitude of different modes, where Eq.~(\eqref{eqwkb}) is of course no longer directly applicable.

With Eq.~\eqref{eqwkb} plugged into Eq.~\eqref{eqgwe} and only keeping the leading order terms in $\epsilon$, the coupled wave equations become a matrix equation of the form
\begin{equation}\label{eqm}
\begin{bmatrix}
\mathcal{H}_1 &  \mathcal{V}_1 \\ \\
 \mathcal{V}_2 &  \mathcal{H}_2 
\end{bmatrix}
\begin{bmatrix}
A_m \\
B_m
\end{bmatrix}
=0\,,
\end{equation}
where the detailed expressions for $\mathcal{H}_1$, $\mathcal{H}_2$, $\mathcal{V}_1$ and $\mathcal{V}_2$ are given in Appendix \ref{sec:FFEqsFlatExp}. The determinant of the above matrix has to be zero for there to be a solution for $A_m$ and $B_m$, which requires that $\mathcal{H}_1 \mathcal{H}_2=\mathcal{V}_1\mathcal{V}_2$ (henceforth refereed to as the ``determinant equation''). After some lengthy but nevertheless straightforward manipulations, we can factorize the determinant equation into the product of the following Hamilton-Jacobi equations
\bea
\frac{\partial S}{\partial t} \pm H_{A, B} =0\,,
\eea
with
\begin{align}\label{eqh}
 H_A &= \sqrt{\frac{m^2}{\rho^2}+\frac{p^2_{\theta}}{r^2}+p_r^2}\,,\nonumber \\
 H_B &= \frac{1}{\left(1+\frac{I(\psi)^2}{4\pi^2 \rho^2}\right)}\left[
- \Omega_F(\psi) \left(\frac{I(\psi)p_z }{2\pi}+m\right)
\pm \right . \nonumber \\
&\left .
\left(p_z - \frac{m I(\psi)}{2\pi \rho^2}\right)
\sqrt{1+ \frac{I(\psi)^2}{4\pi^2 \rho^2}-\rho^2 \Omega_F(\psi)^2 }
\right]\,,
\end{align}
where the ``momenta'' are defined by
\bea
p_C \equiv \frac{\partial S}{\partial C},
\eea
with $C$ being any coordinate. We also note that $p_\phi= m$ and the terms inside the square bracket in $H_B$ is proportional to ${\bf B}^2-{\bf E}^2$, which is a non-negative geometric scalar for the force-free plasma. These two Hamilton-Jacobi equations describe two different sets of modes. The first one is equivalent to
\bea
g^{\mu\nu}k_\mu k_\nu = \eta^{\mu\nu}\partial_{\mu} S \partial_\nu S=0\,,
\eea
which is the eikonal limit of the wave equation for vacuum electromagnetic quasinormal modes. As discussed in Ref.~\cite{1997MNRAS.291..125U} by writing down the local dispersion relation, this family of modes corresponds to the fast-magnetosonic waves in the force-free plasma; it is also consistent with the (eikonlal limit of the) ``trapped mode" classification we found in analyzing the modal structure of the Blandford-Znajek solution \cite{Yang:2014zva}. We emphasize that this agreement is only in the eikonal limit, as the trapped modes differ from the vacuum electromagnetic quasinormal modes for longer wavelengths. 

The propagation of the second family of modes can be determined through
 the Hamiltonian equations of motion. In fact, based on the form of $H_B$ in Eq.~(\ref{eqh}), we can see that
\begin{align}
& \frac{d z}{d t} = \frac{\partial H_B}{\partial p_z} \nonumber \\
& =  \frac{1}{\left(1+\frac{I^2}{4\pi^2 \rho^2}\right)} \left ( - \frac{\Omega_F I}{2 \pi} \pm \sqrt{1+ \frac{I^2}{4\pi^2 \rho^2}-\rho^2 \Omega_F^2 }\right ),\nonumber \\
& \frac{d \phi}{d t} = \frac{\partial H_B}{\partial p_\phi} \nonumber \\
& =  \frac{1}{\left(1+\frac{I^2}{4\pi^2 \rho^2}\right)} \left ( - \Omega_F\mp \frac{I}{2 \pi \rho^2} \sqrt{1+ \frac{I^2}{4\pi^2 \rho^2}-\rho^2 \Omega_F^2 }\right ),
\end{align}
where the $\pm$ sign labels the propagation direction and the equations above suggest that this wave has a null group velocity (as opposed to phase velocity) and freely propagates along the magnetic field lines in the frame of vanishing electric field (i.e. the ``rest-frame" of the magnetic fields). These properties are consistent with those of the Alfv\'en waves in the force-free plasma \cite{1997MNRAS.291..125U}. On the other hand, this family of waves can obviously be seen as the eikonal limit of the ``traveling waves" identified in Ref.~\cite{Yang:2014zva} (in that paper the travelling waves follow the mostly radial monopole-like background magnetic field lines).

Now that we have introduced the necessary WKB techniques for handling coupled wave equations and learned about the mode structure of jets in the flat spacetime, we are now ready to study the wave propagation properties in a slowly-spinning Kerr background, where relativistic effects are non-negligible.

\subsection{Eikonal waves in a slowly-spinning Kerr spacetime}\label{sec43}

With a rotating black hole sourcing the jet, we can still write down the wave equations of the jet perturbations based on Eqs.~\eqref{eq:FFEq1} and \eqref{eq:FFEq2}, subject to the Kerr background:
\begin{align}
ds^2=&-\left(1-\frac{2Mr}{\bar \rho^2}\right)dt^2-\frac{4aMr\sin^2{\theta}}{\bar \rho^2}dt d\phi
\notag \\ &
+\frac{\bar \rho^2}{\Delta}dr^2+\bar \rho^2d\theta^2
\notag \\ & 
+\sin^2{\theta}\left(r^2+a^2+\frac{2Ma^2r\sin^2{\theta}}{\bar \rho^2}\right)d\phi^2\,.
\end{align}
Here $\Delta \equiv r^2-2Mr+a^2$ and $\bar \rho^2=r^2+a^2\cos^2{\theta}$. As discussed in Sec.~\ref{sec2}, the family of jet solutions is only known for $a \ll 1$, in which case $\psi, \Omega_F, I$ are expanded with respect to $a$ and solved order by order (see Eq.~\ref{eqexpanv}). Under this slow rotation limit, we can still Fourier decompose $\alpha$ and $\beta$ as in  Eq.~\eqref{eqfdecom}, and the wave equations have the same form as Eq.~\eqref{eqgwe}, but with the operators now given by
\bea\label{eqgen}
H_i &=& H_i^{(0)} + a H_i^{(1)}+\mathcal{O}(a^2)\,,\nonumber \\
V_i &=& V_i^{(0)} +a V_i^{(1)}+\mathcal{O}(a^2)\,, 
\eea
where
\bea\label{eqh0}
H_1^{(0)}&=& -\frac{\partial^2 }{\partial t^2} + \frac{f \sin^2\theta}{r^2}\frac{\partial^2 }{\partial \cos\theta^2} + f\left( f \frac{\partial }{\partial r}\right)_{,r}\,,\notag \\
H_2^{(0)}&=& -\frac{r-M+M\cos2\theta}{r f} \frac{\partial^2 }{\partial t^2} + f\left( \frac{\sin\theta}{r}\frac{\partial}{\partial \theta} -\cos\theta \frac{\partial }{\partial r}\right)^2  
\notag \\ &&
- \frac{1}{2r}\left[ (2f-1)\cos2\theta -1 \right] \frac{\partial}{\partial r} + \frac{3f-1}{2r^2}\sin2\theta \frac{\partial }{\partial \theta} 
\notag \\ &&
- \frac{m^2}{r^2}\left( \cot^2\theta + f \right)\,,\notag\\
V_1^{(0)}&=&  i f \sin^2\theta m \left(\cot\theta \frac{\partial }{\partial \theta}+ f r  \frac{\partial }{\partial r} \right)\,, \notag \\
V_2^{(0)}&=& \frac{i m}{r^4\sin^2\theta} \left(\cot\theta \frac{\partial}{\partial \theta}+ fr \frac{\partial}{\partial r}\right)  \,
\eea
(notice that $H^{(0)}_{i}$ and $V^{(0)}_{i}$ give the wave equations of the Wald background configuration, with a Schwarzschild black hole and surrounding force-free plasma), 

and 
\begin{widetext}
\bea\label{eqh1}
H_1^{(1)}&=& -\frac{i m}{r^2} \left\{2 r^2 f^2 \frac{\partial \psi_2^{(1)}}{\partial r} \frac{\partial}{\partial r}+2 r^2  \Omega_F^{(1)} \frac{\partial}{\partial t}+2f \frac{\partial \psi_2^{(1)}}{\partial \theta} \frac{\partial}{\partial \theta} 
\notag \right. \\ && \left.
+\left( f^2 r^2 \frac{\partial^2 \psi_2^{(1)}}{\partial r^2}+2 f M \frac{\partial \psi_2^{(1)}}{\partial r}+f \frac{\partial^2 \psi_2^{(1)}}{\partial \theta^2}
-f \cot (\theta ) \frac{\partial \psi_2^{(1)}}{\partial \theta}\right)\right\} 
\,,\\
V_1^{(1)}&=& \frac{1}{2} \sin (\theta ) \left\{
+2 r^2 \sin ^2(\theta )
 \left(f r \sin (\theta ) \frac{\partial \psi_2^{(1)}}{\partial r}+\cos (\theta ) \frac{\partial \psi_2^{(1)}}{\partial \theta}\right)
 \frac{\partial^2}{\partial t^2}
-2 f^2 r^2 \sin ^2(\theta ) \cos (\theta ) \frac{\partial \psi_2^{(1)}}{\partial \theta} \frac{\partial^2}{\partial r^2}
\notag \right. \\ && \left.
-2 f^2 r \sin ^3(\theta ) \frac{\partial \psi_2^{(1)}}{\partial r} \frac{\partial^2}{\partial \theta^2}
+2 f^2 r \sin ^2(\theta )  \left(\sin (\theta ) \frac{\partial \psi_2^{(1)}}{\partial \theta}+r \cos (\theta ) \frac{\partial \psi_2^{(1)}}{\partial r}\right)\frac{\partial^2}{\partial r \partial \theta}
\notag \right. \\ && \left.
-2 f \sin ^3(\theta ) \left(2 M-r^3 \Omega_F^{(1)}\right) \frac{\partial^2}{\partial t \partial r}
 +\frac{\sin (2 \theta ) \sin (\theta )}{r} \left(r^3 \Omega_F^{(1)}-2 M\right) \frac{\partial^2}{\partial t \partial \theta}
\notag \right. \\ && \left. 
\notag \right. \\ && \left.
+\frac{2 \sin (\theta )}{r}  \left[-2M(1+3\cos(\theta)) + 4r^2(r-3M \sin^2(\theta))\Omega_F^{(1)}+r^4(r-2M\sin^2(\theta))\sin^2(\theta) \Omega_F^{(1)'}\right] \frac{\partial}{\partial t}
\notag \right. \\ && \left.
+\sin (\theta )  \left[\sin (2 \theta ) f (8M-5r) \frac{\partial \psi_2^{(1)}}{\partial \theta}
+ 2(3 \cos (2 \theta )+1) f^2 r^2 \frac{\partial \psi_2^{(1)}}{\partial r}
+f^2 r^2 \sin (2 \theta ) 
\frac{\partial^2 \psi_2^{(1)}}{\partial r \partial \theta}
-2 f^2 r \sin ^2(\theta ) \frac{\partial^2 \psi_2^{(1)}}{\partial \theta^2}
\right]
\frac{\partial}{\partial r}
\notag \right. \\ && \left.
-2 \sin ^2(\theta ) \left[f (5r-2M) \cos (\theta ) \frac{\partial \psi_2^{(1)}}{\partial r}-2f r\sin (\theta )  \frac{\partial \psi_2^{(1)}}{\partial \theta}
+f^2 r \left(r \cos (\theta ) \frac{\partial^2 \psi_2^{(1)}}{\partial r^2}-\sin (\theta ) \frac{\partial^2 \psi_2^{(1)}}{\partial r \partial \theta}\right)\right]\frac{\partial}{\partial \theta}
\notag \right. \\ && \left.
+2 f m^2 \left(f r \sin (\theta ) \frac{\partial \psi_2^{(1)}}{\partial r}+\cos (\theta ) \frac{\partial \psi_2^{(1)}}{\partial \theta}\right)
\right\}
\,,
\eea
\bea
H_2^{(1)}&=& -\frac{4 i m M (M \cos (2 \theta )-M+r) }{f r^4}\frac{\partial}{\partial t} \,, \\
V_2^{(1)} &=& \frac{1}{2 r^6} \left\{
\frac{2 r^4}{f}  \left(f r \frac{\partial \psi_2^{(1)}}{\partial r}+\cot (\theta ) \frac{\partial \psi_2^{(1)}}{\partial \theta}\right)\frac{\partial^2}{\partial t^2}
-2 f r^4 \cot (\theta ) \frac{\partial \psi_2^{(1)}}{\partial \theta} 
\frac{\partial^2}{\partial r^2}-2 f r^3 \frac{\partial \psi_2^{(1)}}{\partial r} \frac{\partial^2}{\partial \theta^2}
\notag \right. \\ && \left.
-2 r^2  \left(2 M-r^3 \Omega_F^{(1)}\right) \frac{\partial^2}{\partial t \partial r}
+\frac{2 r \cot (\theta )}{f} \left(r^3 \Omega_F^{(1)}-2 M\right) \frac{\partial^2}{\partial t \partial \theta}
+2 f r^3 \left(\frac{\partial \psi_2^{(1)}}{\partial \theta}+r \cot (\theta ) \frac{\partial \psi_2^{(1)}}{\partial r}\right) \frac{\partial^2}{\partial r \partial \theta}
\notag \right. \\ && \left.
+\frac{4r}{f} \left[M(\cot^2(\theta)-1)+r^2(3M-r\csc^2(\theta))\right]\frac{\partial}{\partial t}
\notag \right. \\ && \left.
+r^2  \left[4 f r \frac{\partial^2\psi_2^{(1)}}{\partial\theta^2}
-4 \cot (\theta )  (5M-2r) \frac{\partial \psi_2^{(1)}}{\partial \theta}
-f r^2  \csc ^2(\theta ) \left(2\sin (2 \theta ) 
\frac{\partial^2\psi_2^{(1)}}{\partial r \partial\theta}
+(3 \cos (2 \theta )+1) \frac{\partial \psi_2^{(1)}}{\partial r}\right)\right] \frac{\partial}{\partial r}
\notag \right. \\ && \left.
+2 r^2 \left[
2 f r^2 \cot (\theta ) \frac{\partial^2 \psi_2^{(1)}}{\partial r^2}-2 f r \frac{\partial^2 \psi_2^{(1)}}{\partial r \partial \theta}+ \cot (\theta ) (2 M+r) \frac{\partial \psi_2^{(1)}}{\partial r}- \frac{\partial \psi_2^{(1)}}{\partial \theta}
\right]\frac{\partial}{\partial \theta}
\notag \right. \\ && \left.
+2 m^2 r^2 \csc ^2(\theta ) \left(f r \frac{\partial \psi_2^{(1)}}{\partial r}+\cot (\theta ) \frac{\partial \psi_2^{(1)}}{\partial \theta}\right)
\right\}\,.
\eea
\end{widetext} 

While the equations are generally coupled in their $\alpha$ and $\beta$ variables, and their $r$ and $\theta$ dependence are not separable, it remains instructive to apply the eikonal approximation, and study the propagation of short-wavelength wave packets.

Using the WKB technique illustrated in Sec.~\ref{sec42}, We first make the ansatz of Eq.~\eqref{eqwkb}, obtain the matrix equation as in Eq.~\eqref{eqm} but with different matrix components (see Appendix \ref{sec:FFEqsKerrExp}), and then solve the determinant equation. After some lengthy calculations we can show that two factor equations for $\partial_t S$ follow naturally from the determinant equation (up to $\mathcal{O}(a)$). The first one reads
\bea \label{eqmag}
 \left (\frac{\partial S}{\partial t}\right)^2+\frac{4 a m M }{r^3}\frac{\partial S}{\partial t}-f^2 p_r^2-\frac{f m^2 \csc ^2\theta }{r^2}-\frac{f p_{\theta }^2}{r^2}=0\,,\nonumber \\
\eea 
which is just the vacuum Teukolsky equation (before separating out the $\theta$ dependence) in the eikonal limit \footnote{See equation 4.7 in \cite{Teukolsky} and keeping only the $\mathcal{O}(a)$ terms}. Therefore, clearly this family of modes are the ``trapped'' modes in the eikonal limit, as we have seen in the flat spacetime case.

\begin{figure}[t,b]\centering
\includegraphics[width=0.45\columnwidth]{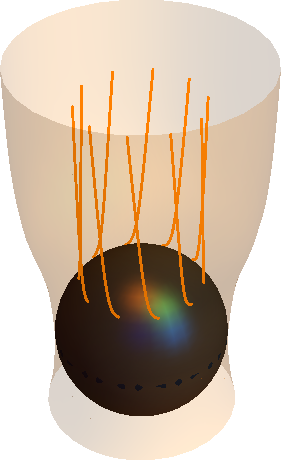}
\caption{A demonstrative example showing the trajectories (orange curves) of the wave packets inside the jet region (bounded by the transparent tube), following the equation of motion as determined by Eq.~\eqref{eqalf}. For this example, the mass of the black hole is chosen to be $1$ and the spin $0.1$. The value of $\Omega_F$ is fixed at $\Omega_H/2$ and $I$ is subsequently given by Eq.~\eqref{eqzna}.}
\label{fig:Traj}
\end{figure}

The second equation contains $I$ and $\Omega_F$, so that this family of modes is affected by the background electromagnetic field, in contrast to the trapped modes. The eikonal wave equation in this case reads 

\begin{align}\label{eqalf}
& I m p_z \left[ f^2 r^2 (M-r)-f^2 M r^2 \cos (2 \theta )\right] \notag \\
&
-\pi F^2 r^4 \frac{\partial S}{\partial t} \sin ^2(\theta ) \left (2  m \Omega_F+\frac{\partial S}{\partial t} \right )
\notag \\ &
+\pi  f^2 F r^3 \sin ^2\theta  p_z^2=0\,,
\end{align}
with
\bea
F \equiv r- 2M \sin^2 \theta,\quad p_z = r \cos \theta  p_r-\sin \theta  p_{\theta }\,.
\eea
We note that Eq.~\eqref{eqalf} suggests that the wave propagates along $z$ and $\phi$ directions. It is thus apparent that the second family of modes are traveling waves, as expected. We present some example trajectories of wave packets following the equation of motion determined by Eq.~\eqref{eqalf} in Fig.~\ref{fig:Traj}.\\

\subsection{Stability of the jet solutions}\label{sec44}

As shown in Sec.\ref{sec3}, given an external magnetic field and a rotating black hole, the relaxed state of the magnetosphere is independent of the initial configuration we chose. On the other hand, we have also shown in Sec.\ref{sec2} that there is a whole family of stationary and axis-symmetric solutions which satisfy the force-free conditions. To reconcile the uniqueness of the steady state with the large available pool of candidates, it is natural to suppose that of all the members of the jet-solution family, only the steady state is in fact stable.

To ascertain if this were the case, we need to examine the stability of the family of solutions, with {\em  modal stability}
being the simplest approach. 
The analysis in Sec.\ref{sec43} and Sec.\ref{sec44} focuses on the wave equations in the eikonal limit, where wave damping or amplification may be ignored as they are slow when compared with wave oscillations. In order to study their stability, we need to go beyond the leading order WKB analysis employed thus far. 
This is a technically arduous task to carry out, and beyond the scope of this work, but we can offer 
some physical arguments to this end. First, since we only impose axis-symmetric initial data in Sec.\ref{sec3}, and such
symmetry is well preserved dynamically, it is sufficient to concentrate on the modes with $m=0$. For these modes,
 Eqs.~\eqref{eqgen}, \eqref{eqh0} and \eqref{eqh1}, suggest that $\alpha$ ``almost" describes the propagation of trapped modes while $\beta$ ``almost" describes the propagation of traveling waves.  This observation follows from the fact that
for the the Schwarzschild background ($a=0$), $\alpha$ and $\beta$ exactly decribe trapped and propagating modes.
For the slowly-spinning-Kerr background, $\alpha$ and $\beta$ fields are are only midly coupled with each other since the off-diagonal terms
(given by $V_{i}$) are of order $\mathcal{O}(a)$, while they principal operators $H_{i}$  change only at order $\mathcal{O}(a^2)$. Consequently the resulting eigen-frequencies can be at most order $\mathcal{O}(a^2)$ away from their Schwarzschild limits \cite{Zach2014}. Now, recall that trapped modes decay in Schwarzschild therefore trapped modes for small values
of $a$ can not suddently become unstable. 
 On the other hand, traveling waves in the Schwarzschild spacetime should have zero damping as their scattering potential is infinitely shallow. Therefore, their counterparts in a slowly-spinning-Kerr spacetime can be unstable with a growth rate $\sim \mathcal{O}(a^2)$ at most. However, this timescale is apparently much longer than the transient period we observe in Sec.\ref{sec3}. The arguments above thus point to modes, if unstable,
only being so in much longer timescales than those observed in the dynamical studies.

Of course, we emphasize that the above arguments are based on modal stability, which is different from linear stability. While our numerical investigation disfavors modal instability as the feature that renders alternative steady state candidates unfavorable, it does not also rule out non-trivial linear instabilities. However, given the broadness of the range of available jet-solutions, our expectation is that this ``selection rule" comes instead from more general mechanisms, such as the physics of current sheets --places where the force-free equations break down as
they signal electrically dominated regions.

\section{Discussion}\label{sec5}
We have presented jet-like solutions of the force-free equations
for spacetimes containing a slowly-spinning black hole.  Our discussion
illustrates that a family of such solutions can be found. Interestingly though,
our numerical
simulations clearly indicate that the configurations tend to change in time and approach
a unique final steady solution, regardless of the initial configuration chosen. We have also carried out further analysis that disfavours modal instability as the triggering mechanism for this dynamical development. Instead, such behavior appears connected to the development of an equatorial current sheet which triggers Alfv\'en
waves propagating through the tube and re-arranging the solution towards the final
one (similar observations have been made in the context of neutron stars~\cite{Gruzinov:2006nn,Ruiz:2012te, Komissarov2004, Komissarov2006, Komissarov2007}).

This would imply that current sheets play a vital role in the dynamical evolution of force-free magnetospheres. It dissipates a portion of the field energy, dynamically feeds back into the field evolution and helps to stabilize the magnetosphere. Therefore a complete study of the stability of different magnetospheric configurations should include current sheets as \emph{dynamical} entities, in addition to the force-free plasma. An analytical treatment of this kind is still lacking, which will be the subject of our future studies. 

We note that one striking consequence of this numerically-observed dynamical behavior is 
that the force-free solution in black hole spacetimes might also obey a sort of ``no-hair theorem'' whereby
the final stable (stationary) solution is uniquely determined by the external field, the black hole mass and its spin.
This is natural to be expected for Einstein+Maxwell black holes, but rather nontrivial with the presence of plasma
and current sheets as we discussed.

As a final comment 
We note that quite differently from tracing the dynamics of individual plasma particles, 
which is usually the topic of study for magnetospheric radiations, 
the ``modes'' investigated in this work represents phonon-like collective motions 
of the entire magnetospheric plasma. 
Such collective modes encode the fundamental parameters of the black hole and the large 
scale structures within the magnetosphere and may be tied to potentially observable phenomena.

\acknowledgements
We thank Peter Goldreich for helpful discussions about collective excitations in the force-free plasma. 
The code employed here (HAD~\cite{had_webpage} which, among 
other things, implemented the force-free equations of motion in
a distributed AMR infrastructure) was mainly  developed by M. Anderson, E. Hirschmann, S.L. Liebling, 
D. Neilsen, C. Palenzuela and LL.
This work was supported by NSERC through a Discovery Grant (to LL) and CIFAR (to LL). FZ would like to thank the Perimeter Institute for Theoretical Physics for hospitality.  
Research at Perimeter
Institute is supported through Industry Canada and by the Province of Ontario
through the Ministry of Research \& Innovation.  
FZ is supported by the National Natural Science Foundation of China grant No.~11443008. 

\appendix
\section{The flat spacetime force-free equations in the eikonal limit \label{sec:FFEqsFlatExp}}

The operators appearing in Eq.~\eqref{eqm} are 
\begin{widetext}
\begin{align}
 \mathcal{H}_1 =& 
-\frac{\sin (\theta )}{r^2} \left\{
r^2\left(r^2 \left(\frac{\partial \psi_2}{\partial r}\right)^2 + \left(\frac{\partial \psi_2}{\partial \theta}\right)^2 + \csc ^2(\theta )\right) \left(\frac{\partial S}{\partial t}\right)^2
\right. \notag \\ & \left. 
2 r^2 p_r \frac{\partial \psi_2}{\partial r} \left(r^2 \Omega _F \frac{\partial S}{\partial t}+p_{\theta } \frac{\partial \psi_2}{\partial \theta}+m \csc ^2(\theta )\right)+2 p_{\theta } \frac{\partial \psi_2}{\partial \theta} \left(r^2 \Omega _F \frac{\partial S}{\partial t}+m \csc ^2(\theta )\right)+2 m r^2 \Omega _F \csc ^2(\theta ) \frac{\partial S}{\partial t}
\right. \notag \\ & \left. 
+r^2 p_r^2 \left(-\left(\frac{\partial \psi_2}{\partial \theta}\right)^2+r^2 \Omega _F^2-\csc ^2(\theta )\right)-p_{\theta }^2 \left(r^2 \left(\frac{\partial \psi_2}{\partial r}\right)^2-r^2 \Omega _F^2+\csc ^2(\theta )\right)
\right. \notag \\ & \left. 
-m^2 \csc ^2(\theta ) \left(r^2 \left(\frac{\partial \psi_2}{\partial r}\right)^2+\left(\frac{\partial \psi_2}{\partial \theta}\right)^2
-r^2 \Omega _F^2 \right)\right\}
\,,
\\
\mathcal{H}_2 =& 
-r^2 \left(-r^2 \sin ^2(\theta ) \left(\frac{\partial S}{\partial t}\right)^2+m^2+\sin ^4(\theta ) p_{\theta }^2+r^2 \sin ^2(\theta ) \cos ^2(\theta ) p_r^2-2 r \sin ^3(\theta ) \cos (\theta ) p_{\theta } p_r\right)
\,,
\\
\mathcal{V}_1 =& 
r \sin (\theta ) p_r \left(r^2 \Omega _F \sin ^2(\theta ) \left(\frac{\partial S}{\partial t}\right)+\sin (\theta ) p_{\theta } \left(\sin (\theta ) \frac{\partial \psi_2}{\partial \theta}+r \cos (\theta ) \frac{\partial \psi_2}{\partial r}\right)+m\right)
\notag \\ & 
+\cos (\theta ) p_{\theta } \left(r^2 \Omega _F \sin ^2(\theta ) \left(\frac{\partial S}{\partial t}\right)+m\right)+\left(r \sin (\theta ) \frac{\partial \psi_2}{\partial r}+\cos (\theta ) \frac{\partial \psi_2}{\partial \theta}\right) \left(r^2 \sin ^2(\theta ) \left(\frac{\partial S}{\partial t}\right)^2-m^2\right)
\notag \\ & 
-r^2 \sin ^2(\theta ) \cos (\theta ) p_r^2 \frac{\partial \psi_2}{\partial \theta}-r \sin ^3(\theta ) p_{\theta }^2 \frac{\partial \psi_2}{\partial r}
\,,
\\
\mathcal{V}_2 =& 
-r p_r \left(r^2 \Omega _F \sin ^2(\theta ) \left(\frac{\partial S}{\partial t}\right)+\sin (\theta ) p_{\theta } \left(\sin (\theta ) \frac{\partial \psi_2}{\partial \theta}+r \cos (\theta ) \frac{\partial \psi_2}{\partial r}\right)+m\right)
\notag \\ & 
-p_{\theta } \left(r^2 \Omega _F \sin (\theta ) \cos (\theta ) \left(\frac{\partial S}{\partial t}\right)+m \cot (\theta )\right)+\left(r \frac{\partial \psi_2}{\partial r}+\cot (\theta ) \frac{\partial \psi_2}{\partial \theta}\right) \left(m^2-r^2 \sin ^2(\theta ) \left(\frac{\partial S}{\partial t}\right)^2\right)
\notag \\ & 
+r^2 \sin (\theta ) \cos (\theta ) p_r^2 \frac{\partial \psi_2}{\partial \theta}+r \sin ^2(\theta ) p_{\theta }^2 \frac{\partial \psi_2}{\partial r}
\,.
\end{align}
\end{widetext}
Note that we have defined the momenta to be
\bea
p_C \equiv \frac{\partial S}{\partial C},
\eea
for any coordinate $C$. We also note that we have the relationship 
\bea
\text{I} =  -2 \pi  r \sin ^2(\theta ) \left(\sin (\theta ) \frac{\partial \psi_2}{\partial \theta}-r \cos (\theta ) \frac{\partial \psi_2}{\partial r}\right)
\eea
from Eq.~\eqref{eq:PolarCurr}, which is utilized when deriving Eq.~\eqref{eqh}.

\section{The Kerr spacetime force-free equations in the eikonal limit \label{sec:FFEqsKerrExp}}

The operators appearing in the Kerr spacetime version of Eq.~\eqref{eqm} are 
\begin{widetext}
\begin{align}
 \mathcal{H}_1 =&
\frac{\csc (\theta )}{f r^2} \left(-2 a f^2 m r^2 p_r \frac{\partial \psi_2^{(1)}}{\partial r}-2 a f m p_{\theta } \frac{\partial \psi_2^{(1)}}{\partial \theta}-r^2 \frac{\partial S}{\partial t} \left(\frac{\partial S}{\partial t}+2 a m \Omega_F^{(1)}\right)+f^2 r^2 p_r^2+f p_{\theta }^2\right)
\,,
\\
\mathcal{H}_2 =& 
-\frac{\sin (\theta )}{4 f} \left\{4 \left[(r-2M \sin^2(\theta)) \left(-4 a m M \sin ^2(\theta ) \frac{\partial S}{\partial t}-r^3 \sin ^2(\theta ) \left(\frac{\partial S}{\partial t}\right)^2+f m^2 r\right)+f^2 r^2 \sin ^4(\theta ) p_{\theta }^2\right]
\right. \notag \\ & \left. 
+f^2 r^4 \sin ^2(2 \theta ) p_r^2-8 f^2 r^3 \sin ^3(\theta ) \cos (\theta ) p_{\theta } p_r\frac{}{}\right\}
\,,
\\
\mathcal{V}_1 =& 
m \left(f r \sin (\theta ) p_r+\cos (\theta ) p_{\theta }\right)
\notag \\ & 
-\frac{a }{2 f r}\left\{r \frac{\partial \psi_2^{(1)}}{\partial \theta} \left(2 \cos (\theta ) \left(f m^2-r^2 \sin ^2(\theta ) \left(\frac{\partial S}{\partial t}\right)^2\right)+f^2 r^2 \sin (2 \theta ) \sin (\theta ) p_r^2-2 f^2 r \sin ^3(\theta ) p_{\theta } p_r\right)
\right. \notag \\ & \left. 
+\sin (\theta ) \left[2 \sin (\theta ) \left(2 M-r^3 \Omega_F^{(1)}\right)  \left(f r \sin (\theta ) p_r+\cos (\theta ) p_{\theta }\right)\frac{\partial S}{\partial t}
\right. \right.  \notag \\ & \left. \left. 
-f r \frac{\partial \psi_2^{(1)}}{\partial r} \left(r^3 \left(\frac{\partial S}{\partial t}\right)^2+r^3 (-\cos (2 \theta )) \left(\frac{\partial S}{\partial t}\right)^2+f r^2 \sin (2 \theta ) p_{\theta } p_r-2 f r \sin ^2(\theta ) p_{\theta }^2+4 m^2 M-2 m^2 r\right)\right]\right\}
\,,\\
\mathcal{V}_2 =& 
-m \left(f r \sin (\theta ) p_r+\cos (\theta ) p_{\theta }\right)
\notag \\ & 
+\frac{a}{2 f r} \left\{r \frac{\partial \psi_2^{(1)}}{\partial \theta} \left(2 \cos (\theta ) \left(f m^2-r^2 \sin ^2(\theta ) \left(\frac{\partial S}{\partial t}\right)^2\right)+f^2 r^2 \sin (2 \theta ) \sin (\theta ) p_r^2-2 f^2 r \sin ^3(\theta ) p_{\theta } p_r\right)
\right. \notag \\ & \left. 
+\sin (\theta ) \left[2 \sin (\theta ) \left(2 M-r^3 \Omega_F^{(1)}\right)  \left(f r \sin (\theta ) p_r+\cos (\theta ) p_{\theta }\right) \frac{\partial S}{\partial t}
\right. \right.  \notag \\ & \left. \left. 
-f r \frac{\partial \psi_2^{(1)}}{\partial r} \left(r^3 \left(\frac{\partial S}{\partial t}\right)^2+r^3 (-\cos (2 \theta )) \left(\frac{\partial S}{\partial t}\right)^2+f r^2 \sin (2 \theta ) p_{\theta } p_r-2 f r \sin ^2(\theta ) p_{\theta }^2+4 m^2 M-2 m^2 r\right)\right]\right\}
\,,
\end{align}
\end{widetext}
where we have expanded the expressions up to $\mathcal{O}(a)$, adopting expanded expressions such as $\Omega_F^{(1)}$ for clarity. Despite the unwieldy appearance of $\mathcal{H}_i$ and $\mathcal{V}_i$, many of their terms cancel out in the determinant equation, which can be written as 
\begin{widetext}
\bea
&&
r^2 \left(\pi  r \sin ^2(\theta ) \left(f^2 p_z^2-r \frac{\partial S}{\partial t} \left(2 a m \Omega_F^{(1)}+\frac{\partial S}{\partial t}\right) F\right)-a I^{(1)} m p_z\right) \left(F \left(r^2 \left(\frac{\partial S}{\partial t}\right)^2-f m^2 \csc ^2(\theta )\right)-f r \left(f p_z^2+q_{\rho }^2\right)\right)
 \notag \\ &=& 
-4 a m M \csc ^2(\theta )  \left[I^{(1)} (M-r) p_z \left(\sin ^2(\theta ) \left(f r \left(f p_z^2+q_{\rho }^2\right)-r^2 \left(\frac{\partial S}{\partial t}\right)^2 F \right)+f m^2 F\right)
\right. \notag \\ && \left.
+\pi  f^2 r^2 \frac{\partial S}{\partial t} \sin ^4(\theta ) p_z^2 F-\pi  r^3 \left(\frac{\partial S}{\partial t}\right)^3 \sin ^4(\theta ) F^2\right]\,.
\label{eq:DetEq}
\eea
\end{widetext}
It is now straight-forward to verify that the product of Eqs.~\eqref{eqmag}, \eqref{eqalf} and $r^2$ agrees with the determinant equation above at $\mathcal{O}(a)$.

\bibliography{References} 

\end{document}